\documentclass [prl,aps,letterpaper,twocolumn,superscriptaddress,amsmath,amssymb,floatfix] {revtex4-1}

\usepackage{mathptmx}

\usepackage[latin9]{inputenc}
\usepackage{amsmath}
\usepackage{amssymb}
\usepackage{graphicx}
\usepackage{esint}
\usepackage{soul}

\usepackage[unicode=true,
 bookmarks=true,bookmarksnumbered=false,bookmarksopen=false,
 breaklinks=false,pdfborder={0 0 1},backref=false,colorlinks=true]
 {hyperref}
\hypersetup{
 linkcolor=magenta, urlcolor=blue, citecolor=blue, pdfstartview={FitH}, hyperfootnotes=false, unicode=true}

\makeatletter

\pdfpageheight\paperheight
\pdfpagewidth\paperwidth

\@ifundefined{textcolor}{}
{%
 \definecolor{BLACK}{gray}{0}
 \definecolor{WHITE}{gray}{1}
 \definecolor{RED}{rgb}{1,0,0}
 \definecolor{GREEN}{rgb}{0,1,0}
 \definecolor{BLUE}{rgb}{0,0,1}
 \definecolor{CYAN}{cmyk}{1,0,0,0}
 \definecolor{MAGENTA}{cmyk}{0,1,0,0}
 \definecolor{YELLOW}{cmyk}{0,0,1,0}
}

\usepackage{xcolor}\usepackage{soul}

\newcommand{\bra}[1]{\ensuremath{\left\langle#1\right|}}
\newcommand{\ket}[1]{\ensuremath{\left|#1\right\rangle}}
\definecolor{blue}{rgb}{0,0,1}
\definecolor{red}{rgb}{1,0,0}
\definecolor{green}{rgb}{0,1,0}
\newcommand{\red}[1]{\textcolor{red}{ #1}}

\makeatother

\begin{document}

%\title{\red{Measurement of the conversion between coherence and quantum correlations with a DQC1 model in a superconducting circuit}}
%\title{\blue{Coherence, discord and  DQC1 model in a superconducting circuit}}
\title{Witnessing quantum resource conversion within deterministic quantum computation using one pure superconducting qubit}

\author{W.~Wang}
\thanks{These two authors contributed equally to this work.}
\affiliation{Center for Quantum Information, Institute for Interdisciplinary Information
Sciences, Tsinghua University, Beijing 100084, China}

\author{J.~Han}
\thanks{These two authors contributed equally to this work.}
\affiliation{Center for Quantum Information, Institute for Interdisciplinary Information
Sciences, Tsinghua University, Beijing 100084, China}

\author{B. Yadin}
\affiliation{Atomic and Laser Physics, Clarendon Laboratory, University of Oxford, Parks Road, Oxford, OX1 3PU, United Kingdom}
\author{Y.~Ma}
\affiliation{Center for Quantum Information, Institute for Interdisciplinary Information
Sciences, Tsinghua University, Beijing 100084, China}
\author{J.~Ma}
\affiliation{Center for Quantum Information, Institute for Interdisciplinary Information
Sciences, Tsinghua University, Beijing 100084, China}
\author{W.~Cai}
\affiliation{Center for Quantum Information, Institute for Interdisciplinary Information
Sciences, Tsinghua University, Beijing 100084, China}
\author{Y.~Xu}
\affiliation{Center for Quantum Information, Institute for Interdisciplinary Information
Sciences, Tsinghua University, Beijing 100084, China}
\author{L.~Hu}
\affiliation{Center for Quantum Information, Institute for Interdisciplinary Information
Sciences, Tsinghua University, Beijing 100084, China}
\author{H.~Wang}
\affiliation{Center for Quantum Information, Institute for Interdisciplinary Information
Sciences, Tsinghua University, Beijing 100084, China}
\author{Y.~P.~Song}
\affiliation{Center for Quantum Information, Institute for Interdisciplinary Information
Sciences, Tsinghua University, Beijing 100084, China}
\author{Mile Gu}
\email{mgu@quantumcomplexity.org}
\affiliation{School of Physical and Mathematical Sciences, Nanyang Technological University, Singapore 639673, Republic of Singapore}
\affiliation{Complexity Institute, Nanyang Technological University, Singapore 639673, Republic of Singapore}
\affiliation{Centre for Quantum Technologies, National University of Singapore, 3 Science Drive 2, Singapore 117543, Republic of Singapore}

\author{L.~Sun}
\email{luyansun@tsinghua.edu.cn}
\affiliation{Center for Quantum Information, Institute for Interdisciplinary Information
Sciences, Tsinghua University, Beijing 100084, China}

%\date{\today}
\begin{abstract}

Deterministic quantum computation with one qubit (DQC1) is iconic in highlighting that exponential quantum speedup may be achieved with negligible entanglement. Its discovery catalyzed heated study of general quantum resources, and various conjectures regarding their role in DQC1's performance advantage. Coherence and discord are prominent candidates, respectively characterizing non-classicality within localized and correlated systems. Here we realize DQC1 within a superconducting system, engineered such that the dynamics of coherence and discord can be tracked throughout its execution. We experimentally confirm that DQC1 acts as a resource converter, consuming coherence to generate discord during its operation. Our results highlight superconducting circuits as a promising platform for both realizing DQC1 and related algorithms, and experimentally characterizing resource dynamics within quantum protocols.

\end{abstract}
\maketitle
\vskip 0.5cm
\newcommand{\comment}[1]{\red{[#1]}}

\narrowtext

Quantum technologies promise to deliver advantages in wide range of information processing tasks from secure communication \cite{gisin2002quantum,scarani2009security}, solving classically intractable problems~\cite{shor1994algorithms,grover1996fast} to the simulation of complex systems~\cite{lloyd1996universal,Georgescu2014,Gu2012}. The historical view held that entanglement enabled this quantum advantage, a quantum resource that plays pivotal roles in many quantum-enhanced protocols \cite{Horodecki2009}. However, this picture is incomplete. For example, universal quantum computation has been shown to be achievable with little entanglement~\cite{VanDenNest2013}. The deterministic quantum computation with one qubit (DQC1) model of computation provides another noteworthy counterpoint~\cite{knill1998power}. The protocol enables potential exponential quantum speedup in evaluating the normalized trace of unitary matrices~\cite{Datta2007PRA,Meyll2015PRA}, and its outputs are hard to sample classically~\cite{Morimae2014PRL,Fujii2018PRL}, yet contains little or no entanglement. This motivated a heated search for alternative explanations regarding its source of quantum advantage~\cite{DattaPRL2008}, and catalyzed the recognition that non-classicality comes in many forms.
 %The rationale being that the better we isolate the peculiar properties of quantum mechanics that enable quantum advantage, the better are which make this possible important both for gaining fundamental insight and devising new applications.

Iconic among such developments was discord \cite{KavanRMP2012}, capturing a more robust form of correlations that can persist in environments where entanglement vanishes. Quantum resources were also proposed to describe non-classical properties to individual systems. This resulted in a framework for quantifying non-classicality within coherent quantum superpositions~\cite{Baumgratz14} that has since undergone extensive study~\cite{Girolami2014observable,Bromley2015Frozen,Yuan15,Cheng2015complementarity,Yao2015quantum,Shao2015fidelity,streltsov2015measuring,ma2016converting,chitambar2016assisted,winter2016operational,Napoli2016Robustness,Yadin2016quantum,Rana2016trace,yuan2016experimental,Radhakrishnan2016PRL,Santos2017}.
Meanwhile, different resources were shown to be convertible into each other, a key example being the use of coherence as a resource for generating quantum correlations \cite{streltsov2015measuring,ma2016converting,chitambar2016assisted,Wu2017}. The manipulation and interplay of these resources is considered a crucial element for understanding the origin of the power of quantum protocols. Indeed, current propositions of how DQC1 gains its operation power include the build-up of discord \cite{DattaPRL2008} and, more recently, the conversion of coherence to quantum correlations~\cite{ma2016converting}.

Here we realize the DQC1 algorithm within a superconducting system with a circuit quantum electrodynamics (QED) architecture~\cite{Wallraff,Paik,Schoelkopf2013}, and monitor the interplay between coherence and discord in the DQC1 model. We implement the algorithm by coherent control and quantum non-demolition (QND) projective measurements on a single pure superconducting qubit dispersively coupled to a harmonic oscillator that can potentially provide a maximally-mixed state with arbitrary dimension. In particular, in our experiment the maximally-mixed state with a dimension of eight is generated with repeated application of Kraus rank-2 channels in an adaptive fashion~\cite{ShenPRB2017} and we perform full joint state tomography on the combined system to characterize the behavior of coherence and discord in the algorithm. Even in the presence of experimental imperfections, we verify that, as theoretically predicted in Ref.~\onlinecite{ma2016converting}, the amount of discord generated during the computation is upper bounded by the consumption of the initial coherence of the system. Our work provides the first experimental characterization of resource conversion dynamics within DQC1, and extends previous experimental realizations of DQC1 in linear optics~\cite{Lanyon2008PRL} and liquid-state nuclear magnetic resonance~\cite{Passante2011PRA} to a new technological medium.

Coherence is taken to mean the superposition of states in some basis set $\{\ket{i}\}$ \cite{Baumgratz14,Streltsov2017,Footnote}. In this work, we focus on one measure: the relative entropy of coherence \cite{Baumgratz14},
\begin{align}
    C(\rho)= S(\rho^{diag})-S(\rho),
\label{Eq:coherence}
\end{align}
where $S(\cdot)$ is the von Neumann entropy and $\rho^{diag}$ is the state obtained by removing the off-diagonal elements of $\rho$ in the reference basis $\{\ket{i}\}$.

In a multipartite system, the quantumness of correlations between subsystems $A_1,\dots,A_n$ may be quantified with the global quantum discord $D(\rho_{A_1,\dots,A_n})$ \cite{Rulli2011pra}. This is defined as the excess of coherence in the global state $\rho_{A_1,\dots,A_n}$ over the local states $\rho_{A_1},\dots,\rho_{A_n}$, minimized over all basis choices for each subsystem (see \cite{Supplement} for details).

\begin{figure}
\includegraphics{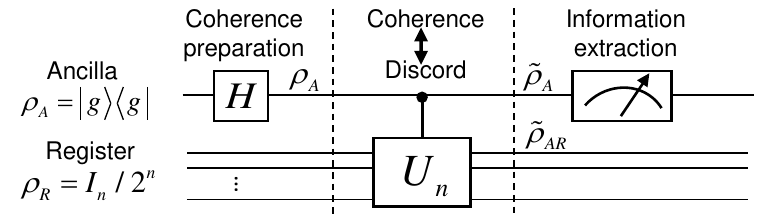} \caption{\textbf{DQC1 model.} Initially the ancilla qubit is prepared in a pure ground state and the register qubits are prepared in a maximally-mixed state. Coherence in the quantum system is then prepared in the ancilla qubit by a Hadamard gate and converted into discord by a controlled operation $U_n$. Measurements of $\langle\sigma_x\rangle$ and $\langle\sigma_y\rangle$ on the ancilla qubit give the real and imaginary parts of $\textrm{Tr}(U_n)/2^n$ respectively.}
\label{fig:DQC1model} \vspace{-6pt}
\end{figure}

The DQC1 protocol is illustrated in Fig.~\ref{fig:DQC1model}. The quantum circuit is fed with $n+1$ qubits, consisting of one pure ancilla qubit and $n$ maximally-mixed register qubits. As noted in Ref.~\onlinecite{ma2016converting}, coherence-to-discord conversion takes place in this algorithm. Specifically, coherence is initially generated in the ancilla qubit by a Hadamard gate. Then a controlled-$U$ gate is performed to correlate the ancilla and the register qubits, and thus create discord between the ancilla and register qubits at the cost of the coherence in the ancilla qubit. Note that the control basis of the gate is taken to be the reference basis for which there is no coherence. This process is encapsulated in the following inequality:
\begin{eqnarray}\label{eq:result}
D(\widetilde{\rho}_{AR})\leq\Delta C(\rho_A),
\end{eqnarray}
where $\widetilde{\rho}_{AR}$ is the joint state of the $n+1$ qubits after the controlled-$U$ gate, $\Delta C(\rho_A)=C(\rho_A)-C(\tilde{\rho}_A)$ is the coherence consumption during the controlled-$U$ gate,  with $\rho_A$ and $\widetilde{\rho}_A$ the states of the ancilla qubit before and after the controlled-$U$ gate, respectively.

\begin{figure}
\includegraphics{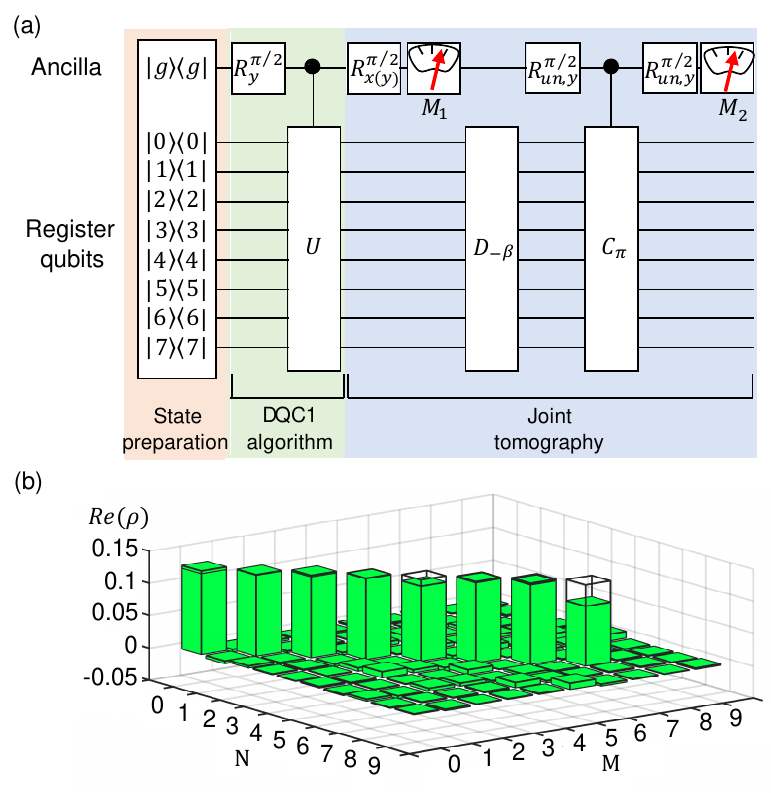} \caption{\textbf{Experimental quantum circuit to measure the conversion between coherence and quantum correlations with a DQC1 model.} (a) The whole process can be divided into three parts: state preparation, DQC1 algorithm, and joint tomography measurement. The maximally-mixed state ($\sum_{k=0}^{7}\ket{k}\bra{k}$) of the initial registers is deterministically generated through a quantum channel construction based on QND measurements of the ancilla and adaptive control of the ancilla-register system. The $R_{y}^{\pi/2}$ operation to create input coherence in the DQC1 algorithm and the following $R_{x(y)}^{\pi/2}$ operations before $M_1$ in the joint tomography are all generated by GRAPE to compensate the extra phases corresponding to different register states due to the dispersive interaction during these gates, such that the gate operations are independent of the states of the register qubits. The controlled-$U$ gate is realized through an appropriate ancilla-register interation time and is used to convert coherence to discord. To characterize the ancilla-register system, joint tomography is performed by correlating the ancilla tomography and subsequent register Wigner tomography. In the joint tomography, the two $\pi/2$ rotations $R_{un,y}^{\pi/2}$ before and after the controlled $\pi$ phase gate $C_{\pi}$ are unconditional gates with a Gaussian envelope of $\sigma=5$~ns. (b) Real part of the reconstructed density matrix (truncated to maximum photon number state $N_{\rm{max}}=9$) of the initial register qubits with a fidelity of 0.977.}
%A $\pi/2$ rotation along the $y$-axis $R_{y}^{\pi/2}$, corresponding to a Hadamard gate on the ancilla qubit, creates input coherence in this circuit. The register states evolve different phase shifts dependent on the ancilla state (a controlled-$U$ gate). To overcome the difficulty of generating a maximally mixed state of the register qubit, we prepare the register qubit in a pure Fock state $\ket{k} (k=0,1,2,3)$ one at a time, run the whole sequence separately, and finally mix the experimental results with equal weight. Joint tomography is performed by correlating the ancilla qubit tomography and subsequent register qubits Wigner tomography (see details in the main text).
\label{fig:sequence} \vspace{-6pt}
\end{figure}

We realize the DQC1 algorithm using a superconducting transmon qubit dispersively coupled to two waveguide cavity resonators~\cite{Paik,Kirchmair,Vlastakis,SunNature,Liu2017}, as shown in Fig.~\ref{fig:sequence}a. The transmon qubit has an energy relaxation time $T_{1}=30$~$\mu$s and a pure dephasing time $T_{\varphi}=120$~$\mu$s. One of the cavities (storage cavity) has a long photon lifetime of $\tau_{s}=143$~$\mu$s. The Fock states in this storage cavity (composing the register qubits) and the transmon qubit (as the ancilla) constitute the bipartite parts of the DQC1 circuit~\cite{KavanRMP2012}. The other short-lived cavity with a photon lifetime $\tau_{r}=44$~ns is used to readout the ancilla qubit. High fidelity and QND single-shot measurements of the ancilla can be achieved with the help of a phase-sensitive Josephson bifurcation amplifier~\cite{Hatridge,Roy,Kamal,Murch}. Each readout measurement throughout our experiment returns a digitized value of the qubit state. The experimental apparatus and readout properties are similar to earlier reports in Refs.~\onlinecite{Hu2019QEC} and \onlinecite{Hu2018channel}.

The ancilla qubit and storage cavity are well described by the dispersive Hamiltonian (omitting small high-order nonlinearities)
\begin{equation}
H/\hbar=\omega_s a^{\dagger}a+\omega_a |e\rangle\langle e|-\chi a^{\dagger}a|e\rangle \langle e|
\label{eq:Hamiltonian}
\end{equation}
where $a^{\dagger}(a)$ is the creation (annihilation) operator of the storage cavity, $|e\rangle$ is the excited state of the ancilla qubit, and $\chi/2\pi=1.90$~MHz is the dispersive interaction strength between the qubit and the storage cavity. This strong dispersive coupling gives rise to the ancilla-register entangling operation, allowing for the controlled-$U$ operation in the DQC1 algorithm. The readout cavity has been neglected since it remains in vacuum unless a measurement is performed.

%The qubit readout is performed by a homodyne detection of the qubit state-dependent cavity response~\cite{Blais2004pra} with the help of a phase-sensitive Josephson bifurcation amplification~\cite{Hatridge,Roy,Kamal,Murch} for a high fidelity and QND single-shot measurement. Each readout measurement throughout our experiment returns a digitized value of the qubit state.

Harmonic oscillators play important roles in quantum information processing ~\cite{Buzek1995,LeghtasPRL2013,Ofek2016,Michael2016,Hu2019QEC} largely due to their infinite dimension and long coherence times. Here we take advantage of these characteristics to use multiple excitations of a harmonic cavity oscillator as register qubit states. In our experiment, we choose the lowest eight Fock states $\{\ket{0}, \ket{1}, ..., \ket{7}\}$ in the cavity field whose computational space is equivalent to that of three register qubits, namely $\{\ket{000},\ket{001},...,\ket{111}\}$ respectively. This gives the DQC1 model in our experiment with $n=3$ register qubits.

%Considering the infinite dimension of the cavity resonator, one could in principle achieve an arbitrarily large DQC1 system.

Our experimental sequence is depicted in Fig.~\ref{fig:sequence}a. The whole process can be divided into three parts: state preparation, DQC1 algorithm, and joint tomography measurement. The state preparation starts with a pure ancilla qubit ground state $|g\rangle\langle g|$ by post-selection. The maximally-mixed state as required by the DQC1 model is deterministically generated through a quantum channel construction based on QND measurements of the ancilla and adaptive control of the ancilla-register system~\cite{ShenPRB2017}. All the adaptive control pulses are numerically calculated with the Gradient Ascent Pulse Engineering (GRAPE) method~\cite{Khaneja2005,deFouquieres2011} and the generated maximally-mixed state has a fidelity of 0.977, whose reconstructed density matrix based on the measured Wigner function is shown in Fig.~\ref{fig:sequence}b. The state fidelity is defined as $F(\rho_{\mathrm{exp}},\rho_{\mathrm{ideal}})=\mathrm{tr}\sqrt{\sqrt{\rho_{\mathrm{ideal}}}\rho_{\mathrm{exp}}\sqrt{\rho_{\mathrm{ideal}}}}$. The protocol for the realization of the maximally-mixed state is shown in \cite{Supplement}. Note that the register qubits do not contribute any coherence to the combined system.

%Our experimental sequence is depicted in Fig.~\ref{fig:sequence}b. The whole process can be divided into three parts: state preparation, DQC1 algorithm, and joint tomography measurement. The state preparation starts with a pure qubit ground state $|g\rangle\langle g|$ by post-selection, however, it is not straightforward in our experiment to generate a maximally mixed state as required by the DQC1 model. For simplicity, we instead initialize the register qubit state to a pure Fock state $\ket{k} (k=0,1,2,3)$ in the cavity field one at a time, run the whole sequence separately, and finally mix the experimental results with equal weight. Note that the register qubits do not contribute any coherence to the combined system. Those Fock states are generated separately by the Gradient Ascent Pulse Engineering method~\cite{Khaneja2005,deFouquieres2011} with fidelities of 0.951, 0.936, 0.929, and 0.924 respectively, based on the corresponding Wigner function.

In the following DQC1 algorithm, coherence preparation is performed by an ancilla qubit operation $R_{y}^{\pi/2}$ corresponding to a $\pi/2$ rotation around the $y$-axis on the Bloch sphere, having the same action here as a Hadamard gate. Note that this $R_{y}^{\pi/2}$ operation and the following $R_{x(y)}^{\pi/2}$ operations before $M_1$ in the joint tomography are also generated by GRAPE to compensate the extra phases during these ``Hadamard" gates, such that the gate operations are independent of the states of the register qubits (see \cite{Supplement}). Consequently, the system is prepared in a product state $\rho_{AR}=(\ket{g}+\ket{e})(\bra{g}+\bra{e})\otimes\sum_{k=0}^{7}\ket{k}\bra{k}$ (ignoring normalization). A conditional cavity phase shift $C_\varphi=\ket{g}\bra{g}\otimes\mathbf{1}+\ket{e}\bra{e}\otimes e^{i\varphi a^{\dagger}a}$ is the mechanism of the controlled-$U$ gate, where $\varphi=\chi t$ is acquired from the free evolution of the dispersive Hamiltonian Eq.~\ref{eq:Hamiltonian} for a time interval $t$. As a result, the controlled-$U$ gate can be described as $\ket{g}\bra{g}\otimes\mathbf{1}+\ket{e}\bra{e}\otimes U$, where
\begin{equation}
U=\left(\begin{array}{cccc}
       1 & 0 & \dots & 0 \\
       0 & e^{i\varphi} & \dots & 0 \\
       \vdots & \vdots & \ddots & \vdots \\
       0 & 0 & \dots & e^{i7\varphi}
     \end{array}
\right)
\label{eq:CUgate}
\end{equation}
in the computational Hilbert space. Note that in the Fock state basis, the controlled-$U$ is an incoherent operation.
After the controlled-$U$ gate, the system evolves to $\tilde{\rho}_{AR}$.

%=\ket{g}\otimes\sum_{k=0}^{7}\ket{k}+\ket{e}\otimes\sum_{k=0}^{7}e^{ik\varphi}\ket{k}$.

To observe the bipartite system, we finally perform a joint measurement of the coupled ancilla-register system with two sequential QND measurements of the ancilla qubit and the register qubits, following a technique similar to that previously demonstrated in Ref.~\onlinecite{VlastakisBell}. As shown in Fig.~\ref{fig:sequence}b, the ancilla qubit detections along one of the three basis vectors $X, Y$, and $Z$ are first performed with or without an appropriate pre-rotation $R_{x(y)}^{\pi/2}$ followed by a $z-$basis measurement $M_1$. These measurements alone can give a full tomography of the ancilla. Then a Wigner tomography of the register qubit is performed by measuring the cavity observable $P(\beta)$ which is a combination of the cavity's displacement operation $D_{-\beta}$ and a parity measurement $P$ of the cavity. The parity measurement is realized in a Ramsey-type experiment of the ancilla qubit, where a conditional cavity $\pi$ phase shift $C_\pi$ is sandwiched in between two unconditional qubit rotations $R_{un,y}^{\pi/2}$ (a Gaussian envelope with $\sigma=5~$ns) followed by another $z-$basis measurement $M_2$~\cite{Lutterbach1997,Bertet2002,Vlastakis,SunNature}. After the qubit tomography measurement $M_1$, the qubit is at a specific known state and the correlation of $M_1$ and $M_2$ determines the parity of the register. Multiplication of the ancilla qubit detection $\sigma_i$ (in the qubit Pauli set $\{I, X, Y, Z\}$) and the register Wigner tomography $W(\beta)=\frac{2}{\pi}\langle P(\beta)\rangle$ shot-by-shot gives the joint Wigner function~\cite{VlastakisBell}, defined as:
\begin{equation}
W_{i}(\beta)=\frac{2}{\pi}\langle\sigma_iP(\beta)\rangle
\label{eq:JointWigner}
\end{equation}

\begin{figure}[tb]
\includegraphics{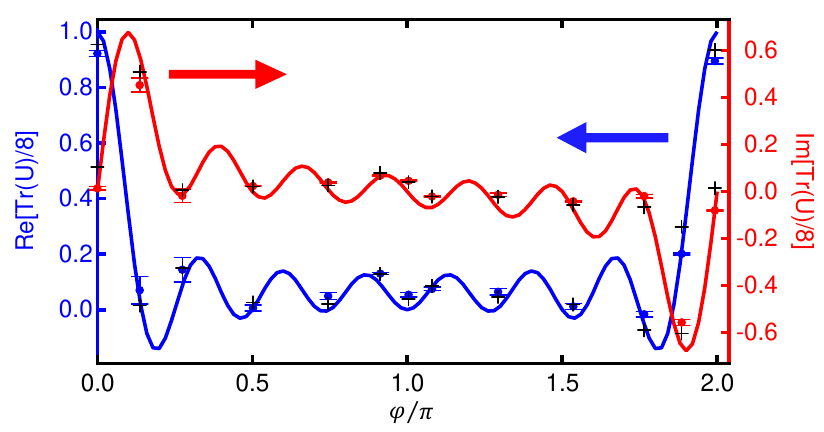} \caption{\textbf{DQC1 algorithm output.} The real and imaginary parts of the normalized trace $\textrm{Tr}(U)/8$ are measured for different $\varphi$. All data (points) are averaged with over $10^7$ measurements and error bars correspond to one standard deviation. The solid lines show theoretical expectations. Crosses present the simulated results including all decoherence channels. Experiment results agree well with both theory and simulation even with the presence of decoherence processes from both ancilla and register qubits.}
%The normalized trace of $U$ is encoded in the ancilla qubit and is recovered from the results of $M_1$ in Fig.~\ref{fig:sequence}b.
\label{fig:DQC1trace} \vspace{-6pt}
\end{figure}

%For any specific phase $\varphi$ in the controlled-$U$ gate, as mentioned above, we generate the initial register qubit state $\ket{k=0,1,2,3}$ one at a time and follow exactly the same circuit in Fig.~\ref{fig:sequence}b individually. Since the Wigner function is linear, this allows to acquire the joint Wigner functions $W_i$ for the case of a mixed initial register state by averaging the four joint Wigner functions $W_{ik}$ corresponding to a pure Fock state $\ket{k}$ as the initial register state. Explicitly, we have:
%\begin{equation}
%W_{i}(\beta)=\frac{1}{4}\sum_{k=0}^{3}W_{ik}(\beta)
%\label{eq:JointWigner}
%\end{equation}

The joint Wigner functions are a complete representation of the combined ancilla-register quantum system. From these functions we reconstruct the combined ancilla-register density matrix $\tilde{\rho}_{AR}$ in a 16-dimensional Hilbert space by a least squares regression using maximum likelihood estimation with the only constraints that the reconstructed density matrix is positive semi-definite with trace equal to one~\cite{Smolin2012,VlastakisBell}. Based on the obtained density matrix $\tilde{\rho}_{AR}$, we can derive the remaining coherence $C(\tilde{\rho}_A)$ and the created discord $D(\tilde{\rho}_{AR})$~\cite{Supplement}, where $\tilde{\rho}_A$ is the partial trace of $\tilde{\rho}_{AR}$ over the registers.

The initial coherence is generated by the ancilla qubit operation $R_{y}^{\pi/2}$ and is first characterized to be $C(\rho_A)=0.894$ by a qubit tomography immediately after this coherent operation, while the ancilla qubit state fidelity is 0.993 (see \cite{Supplement}). The reduction is mainly due to a qubit decay process during the tomography measurement and the imperfection of $R_{y}^{\pi/2}$ in the presence of the maximally-mixed state. This initial coherence built in the ancilla state is then to be consumed in order to correlate the ancilla and the register qubits, and thus is used as a reference for the coherence consumption. We next show the results after applying the controlled-$U$ gate in the DQC1 model. The normalized trace of $U$ is encoded in the ancilla qubit and can be recovered from the result of $M_1$, as shown in Fig.~\ref{fig:DQC1trace}. Although there are decoherence processes with both ancilla and register qubits, the experimental results (dots) agree well with the exact theoretical expectation (lines) and the numerical simulation (crosses) that involves all imperfection channels, suggesting the robustness of the protocol.

\begin{figure}[tb]
\includegraphics{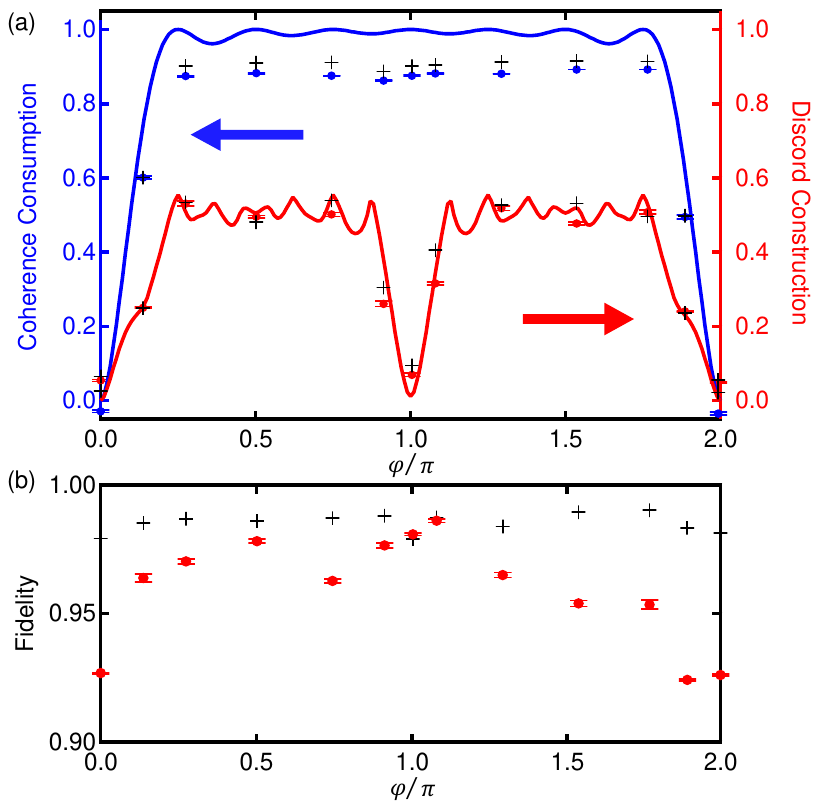}
\caption{\textbf{The coherence consumption $\Delta C(\rho_A)$ and discord production $D(\tilde{\rho}_{AR})$ as a function of phase $\varphi$ in the DQC1 model.} (a) Theoretical expectations are shown with solid lines. Dots give the experimental results.  In the experiment, each point in the joint Wigner functions has been averaged over 3000 single-shot joint ancilla and register measurements, and the standard deviation in $\Delta C(\rho_A)$ and $D(\tilde{\rho}_{AR})$ are estimated by bootstrapping on the measured joint Wigner functions~\cite{VlastakisBell}. Crosses present the simulated data including all decoherence channels with both the ancilla and register qubits. The measured $\Delta C(\rho_A)$ agrees fairly well with the theoretical one with a small gap at the middle plateau. This gap comes from the imperfect generation and measurement of the initial coherence $C(\rho_A)=0.894$ while the ancilla state fidelity is 0.993 to the ideal state $(\ket{g}+\ket{e})/\sqrt{2}$ after the $R_{y}^{\pi/2}$ operation. The measured $D(\tilde{\rho}_{AR})$, in excellent agreement with the theoretical expectation, captures all the important features as in the theoretical curve and is unambiguously lower than $\Delta C(\rho_A)$ as expected. (b) The fidelity of the measured $\tilde{\rho}_{AR}$ compared to the ideal ones with an average of 0.96.}
%The gap comes from the fact that, besides being converted to the discord in the system, the initial coherence of the ancilla qubit inevitably  decays to the surroundings during the algorithm process due to decoherence. The high fidelities, all above 0.92, demonstrate the good control of the quantum system in the whole process.
\label{fig:CoherenceDiscordCurve}
\vspace{-6pt}
\end{figure}

We finally show in Fig.~\ref{fig:CoherenceDiscordCurve} the coherence consumption $\Delta C(\rho_A)$ and discord production $D(\tilde{\rho}_{AR})$ in the DQC1 model as a function of phase $\varphi$ in the controlled-$U$ gate which varies over the range $0\leq \varphi \leq 2\pi$. Theoretical expectations of $\Delta C(\rho_A)$ and $D(\tilde{\rho}_{AR})$ for an ideal system without any decoherence are shown as solid lines for comparison. Both curves are symmetric around $\varphi=\pi$. At $\varphi=\pi$, the controlled-$U$ gate is expected to transform the qubit-register system into a classically correlated state without any discord. At $\varphi=\pi/2, \pi, 3\pi/2$, after tracing out the register state the ancilla qubit is in a maximally-mixed state. In these cases, the original coherence is completely consumed by the controlled-$U$ operation. The gap between the expected solid lines means that, even theoretically, coherence can not be fully converted to discord. The non-monotonic oscillations in the theoretical curves are not fully understood yet and need further investigation.

%To characterize these subtle features in theory, especially for $D(\tilde{\rho}_{AR})$, we replace \red{$5\pi/6$ and $7\pi/6$ with $0.653\pi$ and $1.343\pi$} respectively.

The experimental results for both $\Delta C(\rho_A)$ and $D(\tilde{\rho}_{AR})$ are depicted as dots and are indeed quite symmetric around $\varphi=\pi$, and capture all the important features in the theoretical curves. The measured coherence consumption agrees fairly well with theory. The small gap at the middle plateau comes from the imperfect generation and measurement of the initial coherence $C(\rho_A)=0.894$. The measured produced discord agrees extremely well with the theoretical expectation. All measured $D(\tilde{\rho}_{AR})$ are indeed significantly lower than the measured $\Delta C(\rho_A)$, successfully demonstrating Eq.~\ref{eq:result}. Figure~\ref{fig:CoherenceDiscordCurve}b shows the state fidelity of the ancilla-register system at each $\varphi$ based on the measured $\tilde{\rho}_{AR}$. All fidelities are above 0.92, demonstrating the good control of our system throughout the process.

We note that in Eq.~\ref{eq:Hamiltonian} the higher order corrections to the dispersive term, such as $(\chi'/2)a^{\dagger}a^{\dagger}aa\ket{e}\bra{e}$ ($\chi'$ is typically more than two orders of magnitude smaller than $\chi$), have been ignored because of their small contribution to $\varphi$ compared to the dispersive term in the small photon number limit. However, these higher order non-linear terms in principle allow for generating arbitrary controlled-$U$ by repeated applications of appropriate cavity displacements followed by appropriate waiting, provided the system has enough coherence~\cite{Deutsch1995,LloydPRL1995,Braunstein2005,ZouPRA2016}.

In summary, we have experimentally demonstrated and quantified quantum resource conversion in the DQC1 model. We show that coherence is converted into the quantum discord that is considered as the resource in DQC1~\cite{DattaPRL2008}. The produced discord is unambiguously demonstrated to be lower than the coherence consumption. This experiment reveals the potential of superconducting circuits as a versatile platform for investigating and even deepening our understanding of resource dynamics in quantum information.

%This is accomplished based on a superconducting circuit QED architecture with an intrinsic dispersive coupling between a superconducting transmon qubit and a waveguide cavity.

A natural extension of the present work is to chain multiple DQC1 circuits together. Provided the register qubits are not reset between iterations, the resulting circuit enables a variant of Shor's algorithm~\cite{Parker2000,Martin2012NP}. Here, each iteration converts an additional bit of coherence into quantum correlations, enabling study of resource conversion dynamics within the iconic quantum factoring protocol. Our architecture is also suitable for the ``power of one pure qumode" protocol -- in which the control qubit in DQC1 is replaced with a continuous variable mode to form a hybrid model of computation that combines discrete and continuous variables~\cite{nana2016}. Its realization could enable hybrid factoring algorithms, and aid the study of how continuous and discrete notions of non-classicality interact. Each of these developments would provide a promising experimental platform for studying quantum resource dynamics within more complex settings.

\clearpage{}


\begin{thebibliography}{71}
\bibitem [{Foo()}]{Footnote}%
  \BibitemOpen
  \href@noop {} {\bibinfo  {journal} {{The reference basis $\{\ket{i}\}$
  typically corresponds to be a basis that is considered classical for physical
  reasons. Examples include the energy eigenbasis, or the basis a system
  naturally decoheres to when left in contact with the environment}}\
  }\BibitemShut {NoStop}%
\end{thebibliography}
\end{document}

\title{Supplementary Materials for ``Witnessing quantum resource conversion within deterministic quantum computation using one pure superconducting qubit"}

\author{W.~Wang}
\thanks{These two authors contributed equally to this work.}
\affiliation{Center for Quantum Information, Institute for Interdisciplinary Information
Sciences, Tsinghua University, Beijing 100084, China}

\author{J.~Han}
\thanks{These two authors contributed equally to this work.}
\affiliation{Center for Quantum Information, Institute for Interdisciplinary Information
Sciences, Tsinghua University, Beijing 100084, China}

\author{B. Yadin}
\affiliation{Atomic and Laser Physics, Clarendon Laboratory, University of Oxford, Parks Road, Oxford, OX1 3PU, United Kingdom}
\author{Y.~Ma}
\affiliation{Center for Quantum Information, Institute for Interdisciplinary Information
Sciences, Tsinghua University, Beijing 100084, China}
\author{J.~Ma}
\affiliation{Center for Quantum Information, Institute for Interdisciplinary Information
Sciences, Tsinghua University, Beijing 100084, China}
\author{W.~Cai}
\affiliation{Center for Quantum Information, Institute for Interdisciplinary Information
Sciences, Tsinghua University, Beijing 100084, China}
\author{Y.~Xu}
\affiliation{Center for Quantum Information, Institute for Interdisciplinary Information
Sciences, Tsinghua University, Beijing 100084, China}
\author{L.~Hu}
\affiliation{Center for Quantum Information, Institute for Interdisciplinary Information
Sciences, Tsinghua University, Beijing 100084, China}
\author{H.~Wang}
\affiliation{Center for Quantum Information, Institute for Interdisciplinary Information
Sciences, Tsinghua University, Beijing 100084, China}
\author{Y.~P.~Song}
\affiliation{Center for Quantum Information, Institute for Interdisciplinary Information
Sciences, Tsinghua University, Beijing 100084, China}
\author{Mile Gu}
\email{mgu@quantumcomplexity.org}
\affiliation{School of Physical and Mathematical Sciences, Nanyang Technological University, Singapore 639673, Republic of Singapore}
\affiliation{Complexity Institute, Nanyang Technological University, Singapore 639673, Republic of Singapore}
\affiliation{Centre for Quantum Technologies, National University of Singapore, 3 Science Drive 2, Singapore 117543, Republic of Singapore}

\author{L.~Sun}
\email{luyansun@tsinghua.edu.cn}
\affiliation{Center for Quantum Information, Institute for Interdisciplinary Information
Sciences, Tsinghua University, Beijing 100084, China}

\maketitle

\section{Theoretical preliminaries}
Here we present theoretical preliminaries for calculations of coherence consumption and discord production. Coherence is taken here to mean the superposition of states in some preferred basis set $\{\ket{i}\}$. There is often a natural physical basis to consider -- here, it will be the energy basis of the superconducting qubit. The  resource theory of coherence~\cite{Baumgratz14} provides criteria for determining valid measures of coherence. One first defines the set of states with no coherence $\mathcal{I}$, or incoherent states, to be all states of the form $\sigma=\sum_ip_i\ketbra{i}{i}$. A good coherence measure must vanish exactly on this set. Next, one defines incoherent operations to be those quantum operations having a set of Kraus operators $\{K_i\}$ with the property $K_i \mathcal{I} K_i^\dagger \in \mathcal{I}$. This condition says that an incoherent operation is never able to create coherence from an incoherent state. The next criterion for a valid coherence measure is then that it cannot increase under an incoherent operation. A number of interesting measures satisfying these criteria have been found~\cite{Streltsov2017}. We use the relative entropy of coherence \cite{Baumgratz14}, defined as
\begin{align}
    C(\rho) & := \min_{\sigma \in \mathcal{I}} \; S(\rho || \sigma) \\
     &= S(\rho^{diag})-S(\rho),
\label{Eq:coherence}
\end{align}
where $S(\rho) = -\tr \rho \log \rho$ is the von Neumann entropy, $S(\rho||\sigma) = -S(\rho) - \tr \rho \log \sigma$ is the relative entropy, and $\rho^{diag}$ is the state obtained by removing the off-diagonal elements of $\rho$ in the reference basis $\{\ket{i}\}$.

Quantum discord has a number of characterizations, the first historically being the gap between the total correlations and the classical correlations accessible from local measurements \cite{henderson2001classical,ollivier2001quantum}. In a system partitioned into $n$ subsystems $\{A_1,A_2,\dots,A_n\}$, states with vanishing discord are called classically correlated. They take the form $\sum_{k_{1...n}}p_{k_{1...n}}\ketbra{k_{1...n}}{k_{1...n}}$, where $\{p_{k_{1...n}}\}\ge 0$, $\sum_{k_{1...n}}p_{k_{1...n}}=1$ and $\{\ket{k_{1...n}}\}=\ket{k_1}\otimes...\otimes\ket{k_n}$ is an arbitrary product basis. Classically correlated states are separable (not entangled), but not every separable state is classically correlated. There are many proposed measures of discord~\cite{KavanRMP2012,Adesso2016}; we focus on one measure which treats all subsystems equally, named global discord~\cite{rulli2011pra}. This is defined by
\begin{equation}
\begin{aligned}
D(\rho_{A_1,\ldots,A_n})&=&\min_{\Phi^i} \; S(\rho_{A_1,\ldots,A_n}\|\Phi^i(\rho_{A_1,\ldots,A_n}))\\
&&-\sum_k S(\rho_{A_k}\|\Phi_{A_k}^i(\rho_{A_k}))
\label{eq:discord}
\end{aligned}
\end{equation}
where $\Phi^i=\otimes^n_{j=1}\Phi_{A_{j}}^i$, $\Phi_{A_k}^i(\rho_{A_k})=\sum_i|i_k\rangle\langle i_k|\rho_{A_k}|i_k\rangle\langle i_k|$ is the dephasing operation, and the minimization is over all dephasing basis choices.

\begin{figure}[htb]
\includegraphics{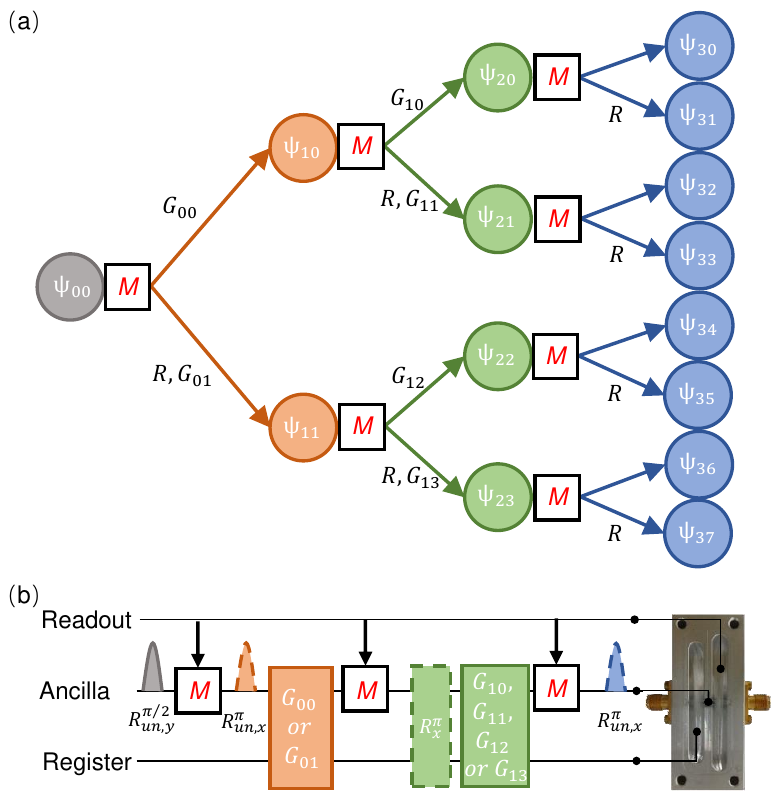}
\caption{\textbf{Preparation of the initial registers in the maximally-mixed state.} (a) Binary tree representation with a depth $L=3$. $M$ represents the QND measurement of the ancilla qubit. States $\ket{\psi}$ and adaptive controls $G$ are presented in Table~\ref{Table:BinaryTreeProcess}. All the adaptive control $G$'s are achieved through numerical optimization with the GRAPE method. $R$ represents a reset process of the ancilla qubit. By implementing three layers of the adaptive operations, we can obtain the maximally-mixed state $\sum_{k=0}^{7}\ket{k}\bra{k}$. (b) The experimental sequence for preparing the initial register state and optical image of the device with a transmon qubit, as the ancilla qubit, dispersively coupled to two three-dimensional cavities. We first prepare the ancilla-register system in $\ket{\psi_{00}}=(\ket{g}+\ket{e})/\sqrt{2}\otimes\ket{0}$ from $\ket{g}\otimes\ket{0}$ through an unconditional qubit rotation $R_{un,y}^{\pi/2}$ (a Gaussian envelope with $\sigma=5$~ns). We then repeatedly measure the ancilla qubit (three layers) along the $z$-axis. Based on the measurement results, proper unitary operations $G_{ij}$ are applied to change the system state to $\ket{\psi_{i+1,j}}$. If the ancilla is measured in the excited state $\ket{e}$, a reset process ``$R$" first flips the qubit to the ground state $\ket{g}$ before the $G$ gate. The reset gates after the first and the third measurements are unconditional $\pi$ pulses $R_{un,x}^{\pi}$. The one after the second measurement is a numerically optimized pulse $R_{x}^{\pi}$ with the GRAPE method, independent of the register state.}
\label{fig:binarytree} \vspace{-6pt}
\end{figure}

\section{Realization of the maximally-mixed state of the register qubits}
As pointed out by Lloyd and Viola~\cite{LloydPRA2001}, universal dynamical control of a quantum system can be achieved if one can perform quantum non-demolition (QND) measurements of the system and feedback the measurement results via coherent control. Shen et al.~\cite{shenPRB2017} further prove that quantum channel simulation of arbitrary dimension can be efficiently realized with a single ancilla qubit and adaptive control in a binary-tree configuration~\cite{AnderssonPRA2008}.

Here, we adopt such a method by repeatedly applying Kraus rank-2 channels with real-time feedback control to generate the maximally-mixed state $\sum_{k=0}^{7}\ket{k}\bra{k}$. Figure~\ref{fig:binarytree} shows the three-layer binary tree representation and the experimental sequence of this protocol. All the adaptive control operations are achieved through numerical optimization with the Gradient Ascent Pulse Engineering (GRAPE) method~\cite{Khaneja2005,deFouquieres2011}, an optimization algorithm designed to numerically find pulses that most accurately realize a unitary operation. The created maximally-mixed state is characterized by the typical Wigner tomography and the reconstructed density matrix is shown in Fig.~2b of the main text with a high state fidelity of 0.977.

%We first prepare a state of system as $(\ket{g}+\ket{e})\ket{0}$, then measure the ancillary qubit along z-basis. if the measurement result is $\ket{g}$, then implement $G_{00}$ to the system, if the measurement result is $\ket{e}$, reset the ancillary qubit first and then implement $G_{01}$ to the system. Next, measure the ancillary qubit again, if the results of first two measurements are $\ket{g}$ and $\ket{g}$, implement $G_{10}$ to the system, if the results are $\ket{g}$ and $\ket{e}$, reset the ancillary qubit and implement $G_{11}$ to the system, if the results are $\ket{e}$ and $\ket{g}$, implement $G_{12}$ to the system, and if the results are $\ket{e}$ and $\ket{e}$, then reset the ancillary qubit and implement $G_{13}$ to the system. Through three layers of similar operations, we can obtain the maximally mixed state $\sum_{k=0}^{7}\ket{k}\bra{k}$. The GRAPE pulses and states in this process are shown in TABLE \Rmnum{1}, the states in the table are not normalized for convenience.

\begin{table}[b]
\caption{The states $\ket{\psi}$ and the adaptive control pulses $G$ in the binary-tree realization of the maximally-mixed state. All $G$'s are achieved through the GRAPE method. The states are not normalized for simplicity.}
\begin{center}
\begin{tabular*}{8cm}{@{\extracolsep{\fill}}lccccc}
\hline\hline
$\psi_{00}$ & $(\ket{g}+\ket{e})\otimes\ket{0}$\\
$\psi_{10}$ & $\ket{g}\otimes(\ket{0}+\ket{7})+ \ket{e}\otimes(\ket{2}+\ket{5})$\\
$\psi_{11}$ & $\ket{g}\otimes(\ket{1}+\ket{6})+ \ket{e}(\ket{3}+\ket{4})$\\
$\psi_{20}$ & $\ket{g}\otimes\ket{7}+\ket{e}\otimes\ket{0}$\\
$\psi_{21}$ & $\ket{g}\otimes\ket{5}+\ket{e}\otimes\ket{2}$\\
$\psi_{22}$ & $\ket{g}\otimes\ket{6}+\ket{e}\otimes\ket{1}$\\
$\psi_{23}$ & $\ket{g}\otimes\ket{4}+\ket{e}\otimes\ket{3}$\\
$\psi_{30}$ & $\ket{g}\otimes\ket{7}$\\
$\psi_{31}$ & $\ket{g}\otimes\ket{0}$\\
$\psi_{32}$ & $\ket{g}\otimes\ket{5}$\\
$\psi_{33}$ & $\ket{g}\otimes\ket{2}$\\
$\psi_{34}$ & $\ket{g}\otimes\ket{6}$\\
$\psi_{35}$ & $\ket{g}\otimes\ket{1}$\\
$\psi_{36}$ & $\ket{g}\otimes\ket{4}$\\
$\psi_{37}$ & $\ket{g}\otimes\ket{3}$\\
\hline
$G_{00}$ & $\ket{g}\otimes\ket{0} \rightarrow \ket{g}\otimes(\ket{0}+\ket{7})+ \ket{e}\otimes(\ket{2}+\ket{5})$\\
$G_{01}$ & $\ket{g}\otimes\ket{0} \rightarrow \ket{g}\otimes(\ket{1}+\ket{6})+ \ket{e}\otimes(\ket{3}+\ket{4})$\\
$G_{10}$ & $\ket{g}\otimes(\ket{0}+\ket{7}) \rightarrow \ket{g}\otimes\ket{7}+\ket{e}\otimes\ket{0}$\\
$G_{11}$ & $\ket{g}\otimes(\ket{2}+\ket{5}) \rightarrow \ket{g}\otimes\ket{5}+\ket{e}\otimes\ket{2}$\\
$G_{12}$ & $\ket{g}\otimes(\ket{1}+\ket{6}) \rightarrow \ket{g}\otimes\ket{6}+\ket{e}\otimes\ket{1}$\\
$G_{13}$ & $\ket{g}\otimes(\ket{3}+\ket{4}) \rightarrow \ket{g}\otimes\ket{4}+\ket{e}\otimes\ket{3}$\\
\hline\hline
\end{tabular*}
\end{center}
\label{Table:BinaryTreeProcess}
\end{table}

\begin{figure}[tbh]
\includegraphics{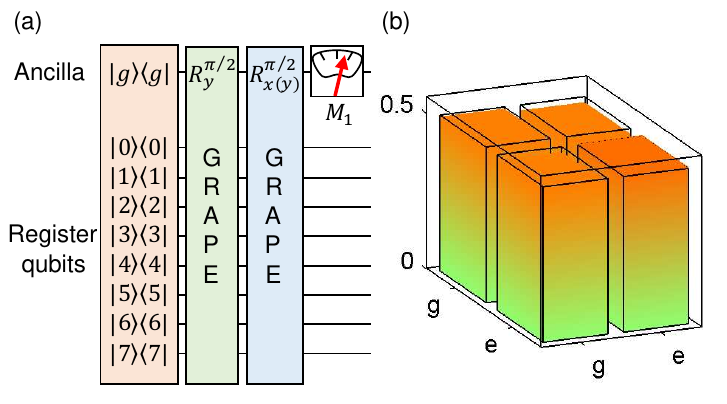} \caption{\textbf{Input coherence calibration.} (a) To calibrate the input coherence, we first prepare the maximally-mixed state $\sum_{k=0}^{7}\ket{k}\bra{k}$ in the storage cavity (the register qubits), then rotate the ancilla qubit to $(\ket{g}+\ket{e})/\sqrt{2}$ with a pulse numerically optimized with the GRAPE method and independent of photon numbers in the cavity. Finally, we measure the ancilla qubit along $X$, $Y$, and $Z$-axis through pre-rotations $R_{x(y)}^{\pi/2}$ of the ancilla followed by a $z$-basis measurement. These pre-rotations are also realized by the GRAPE methods and insensitive to the state of the cavity. (b) The experimental density matrix of the ancilla qubit after being prepared in $(\ket{g}+\ket{e})/\sqrt{2}$ for the initial coherence. The state fidelity is 0.993 and its density matrix contributes 0.894 of the relative entropy of coherence. The solid boxes represent theoretical values.}
\label{fig:inputcoherence} \vspace{-6pt}
\end{figure}

\begin{figure}[tbh]
\includegraphics{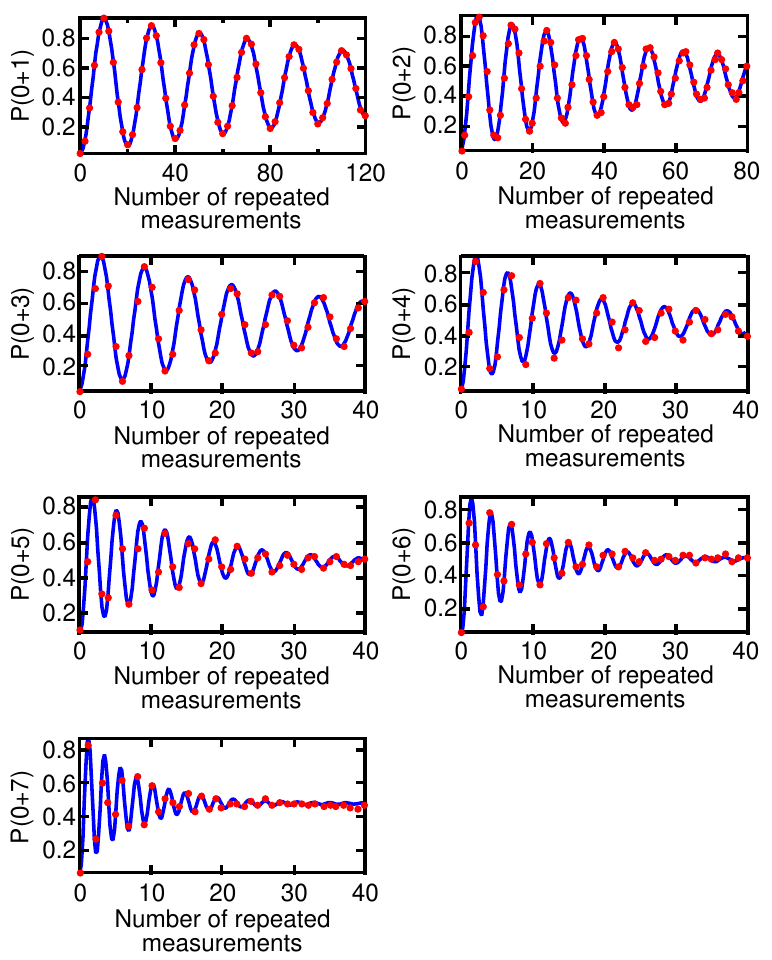} \caption{\textbf{Calibration of the measurement-induced phases.} The registers are first prepared in $(\ket{0}+\ket{1})/\sqrt{2}$, $(\ket{0}+\ket{2})/\sqrt{2}$, ..., $(\ket{0}+\ket{7})/\sqrt{2}$ respectively, and then the probabilities $p(0+N)$ of the registers along the corresponding basis of $(\ket{0}+\ket{1})/\sqrt{2}$, $(\ket{0}+\ket{2})/\sqrt{2}$,..., $(\ket{0}+\ket{7})/\sqrt{2}$ are measured after repeated implementations of measurement operations. Dots are experimental data and solid lines are fit with damped sinusoidal functions.}
\label{fig:RIP} \vspace{-6pt}
\end{figure}

\section{Manipulation of the ancillary qubit independent of the register state and calibration of the input coherence}
In the DQC1 algorithm, initially the ancilla qubit is prepared in a pure ground state and the register qubits are prepared in a maximally-mixed state. Coherence in the quantum system is prepared in the ancilla qubit by a Hadamard gate and then converted into discord by the following controlled operation. To have this algorithm work properly, it is critical that the Hadamard gate on the ancilla qubit is independent of the register state. As the registers are in the maximally-mixed state, photon numbers in the storage cavity are variable. Therefore, a conventional rotation pulse (with a Gaussian envelope) on the ancilla qubit will no longer work because different photon numbers induce different phases on the ancilla qubit -- as the duration required to execute the pulse is non-negligible in the context of the coupling strength between the ancilla and the cavity mode. To mitigate these phase errors, we instead use the GRAPE method to realize the Hadamard gate, which does not depend on the number of photons in the cavity mode. The following ancilla rotations (the operations before $M_1$ in the joint tomography process, Fig.~2a of the main text) also need to be insensitive to the register state and are realized in the same way.

Figure~\ref{fig:inputcoherence}a shows the experimental circuit to calibrate the input coherence, which also confirms the insensitivity of the ancilla qubit rotations to the photon numbers in the storage cavity. The experimental results for the generated $(\ket{g}+\ket{e})/\sqrt{2}$ when the registers are in the maximally-mixed state are shown in Fig.~\ref{fig:inputcoherence}b with a state fidelity of 0.993. The corresponding density matrix contributes 0.894 of the relative entropy of coherence based on Eq.~1 of the main text.

\begin{table}[t]
\caption{The measurement-induced phase for each photon number state in the cavity.}
\begin{center}
\begin{tabular*}{8cm}{@{\extracolsep{\fill}}lccccc}
\hline\hline
photon number state & measurement-induced phase (rad)\\
\hline
$\ket{1}$ & 0.31\\
$\ket{2}$ & 0.65\\
$\ket{3}$ & 1.03\\
$\ket{4}$ & 1.43\\
$\ket{5}$ & 1.85\\
$\ket{6}$ & 2.30\\
$\ket{7}$ & 2.78\\
\hline\hline
\end{tabular*}
\end{center}
\label{Table:MeasInducedPhase}
\end{table}

\section{Calibration and compensation of the measurement-induced phases}
Due to the coupling between the readout cavity and the storage cavity, each measurement induces a phase to the register state. To calibrate this measurement-induced phase, we prepare the registers in $(\ket{0}+\ket{1})/\sqrt{2}$, $(\ket{0}+\ket{2})/\sqrt{2}$,..., $(\ket{0}+\ket{7})/\sqrt{2}$ respectively, implement the measurement operations repeatedly, and finally measure the probabilities $p(0+N)$ of the registers along the corresponding basis of $(\ket{0}+\ket{1})/\sqrt{2}$, $(\ket{0}+\ket{2})/\sqrt{2}$, ..., $(\ket{0}+\ket{7})/\sqrt{2}$ (Ramsey-type experiments). The experimental results are shown in Fig.~\ref{fig:RIP}. Fitting the curves with damped sinusoidal functions gives the induced phases by each measurement for different photon number states (TABLE~\ref{Table:MeasInducedPhase}). To generate the initial register state accurately, we compensate these extra phases induced by the measurement operation in the experiment when optimizing the GRAPE pulses.

\bibliographystyle{Zou}
%\bibliography{bibliography}
%Control: production of eprint (-1) disabled
%

\clearpage{}